# Topological links and knots of speckled light mediated by coherence singularities

Zhuoyi Wang[1,5], Xingyuan Lu[1,5], Zhigang Chen[2✉], Yangjian Cai[3,4✉] and Chengliang Zhao[1✉]

[1] School of Physical Science and Technology, Jiangsu Key Laboratory of Frontier Material Physics and Devices & Suzhou Key Laboratory of Intelligent Photoelectric Perception, Soochow University, Suzhou 215006, China.

[2] The MOE Key Laboratory of Weak-Light Nonlinear Photonics, TEDA Applied Physics Institute and School of Physics, Nankai University, Tianjin 300457, China.

[3] Shandong Provincial Engineering and Technical Center of Light Manipulations & Shandong Provincial Key Laboratory of Optics and Photonic Device, School of Physics and Electronics, Shandong Normal University, Jinan 250358, China.

[4] Joint Research Center of Light Manipulation Science and Photonic Integrated Chip, East China Normal University, Shanghai 200241, China.

[5] These authors contributed equally.

✉e-mail: zgchen@nankai.edu.cn (Z. Chen); yangjiancai@sdnu.edu.cn (Y. Cai); zhaochengliang@suda.edu.cn (C. Zhao).

**Abstract**. Links and knots are exotic topological structures that have garnered significant interest across multiple branches of natural sciences. Coherent links and knots, such as those constructed by phase or polarization singularities of coherent light, have been observed in various three-dimensional optical settings. However, incoherent links and knots — knotted or connected lines of coherence singularities — arise from a fundamentally different concept. They are "hidden" in the statistic properties of a randomly fluctuating field, making their presence often elusive or undetectable. Here, we theoretically construct and experimentally demonstrate such topological entities of incoherent light. By leveraging a state-of-the-art incoherent modal-decomposition scheme, we unveil incoherent topological structures from fluctuating light speckles, including Hopf links and Trefoil knots of coherence singularities that are robust against coherence and intensity fluctuations. Our work is applicable to diverse wave systems where incoherence or practical coherence is prevalent, and may pave the way for design and implementation of statistically-shaped topological structures for various applications such as high-dimensional optical information encoding and optical communications.

## Introduction

In recent years, the investigation of topological structures has attracted significant attention due to their fundamental interest and potential applications. Links and knots represent typical topological structures, defined mathematically as closed curves in three-dimensional (3D) space. appearing as interlinked strings and chains [1-3]. These structures have been observed in various physical systems, including classical and quantum fields [4-6], fluid dynamics [7], crystals [8-10], and plasmas [11], playing a prominent role in physical sciences [12-15]. In particular, as complex 3D optical topological structures, links and knots of optical fields have sparked extensive experimental and theoretical research interests, leading to the exploration of exotic dynamics of knotted vortices [16-21].

With coherent light fields, optical knots composed of phase singularities were first theoretically reported by Berry and Dennis in 2001 [22]. Subsequently, Leach et al. experimentally constructed such knots based on linear superposition of fully coherent Laguerre–Gaussian (LG) beams with optimized intensities and phases [23-26]. Following this, the singularity line of the polarization structure was proposed, creating a space-varying polarized beam with singularities that undergo knotted trajectories during free-space propagation [27-29]. In the generation of optical links or knots, fully coherent beams have been predominantly considered and employed [30-34]. However, it is worth noting that apart from phase and polarization, coherence is also one of the fundamental characteristics of a light field. Indeed, coherence singularities have been proposed and used to characterize the point pairs with zero degrees of coherence [35-37], even though their intensities are not necessarily zero. Yet, to our knowledge, the topological nature of coherence singularities arising from incoherent light fields has never been explored.

In this study, we introduce a new class of 3D topological entities termed incoherent links and knots. These structures represent knotted or connected lines of coherence singularities that are hidden within the coherence structures of incoherent light fields. Through experimental investigations using a modal decomposition measurement approach [38,39], we examine a series of fluctuating speckled light patterns and uncover incoherent Hopf links and Trefoil knots. Our findings indicate that while the phase singularities within these fields exhibit variability and may wash out completely, the coherence singularities remain robust, forming incoherent links or knots observed in the evolution of coherence structure during beam propagation. The discovery of these incoherent topological structures may inspire further interest and investigation in turbulence-resistant 3D topological beam structures, incoherent light-matter interactions, and high-dimensional optical information coding.

**Results**

Within a coherent light field, a point exhibiting zero intensity and an indeterminate phase is referred to as a phase singularity. In the context of the polarization dimension, a point characterized by an undefined polarization vector is referred to as a polarization singularity. In contrast, coherence singularities are "hidden" within the statistical properties of an incoherent light field and do not correspond to conventional vortex singularities or intensity zeros. Figure 1a illustrates a schematic diagram of coherence singularities in incoherent light, specifically depicted as fluctuating speckled light. Due to statistical averaging, the intensity of speckled light exhibits a uniform distribution. The degree of coherence $|\mu|$, as a second-order correlation statistic of speckled light, is characterized by the interference between point pairs $(\boldsymbol{\rho}, \boldsymbol{\rho}_0)$ (Fig. 1a). The interference fringes of intensity patterns between different paired points are shown in Fig. 1(a). $\boldsymbol{\rho}$ represents the two-dimensional coordinates in the *x-y* plane at a fixed propagation distance *z*. Specifically, a zero degree of coherence ($|\mu(\boldsymbol{\rho}_b, \boldsymbol{\rho}_0)| = 0$) indicates a coherence singularity at $\boldsymbol{\rho}_b$ (for a fixed reference point at $\boldsymbol{\rho}_0$) [35]. By connecting multiple stable coherence singularities in speckled light, incoherent Hopf links can be generated. As shown in Fig. 1b, blue and red are used to distinguish between different rings. In contrast, due to the fluctuating intensity, amplitude, and phase of speckled light (Fig. 1c, d), no stable phase singularities can be observed.

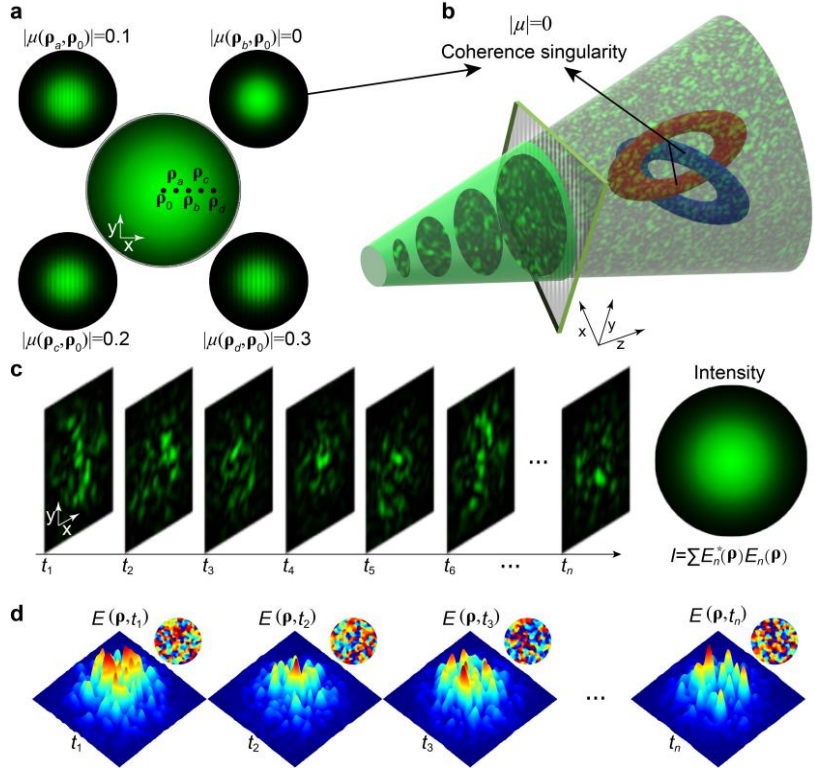

**Fig. 1 Schematic diagram illustrating coherence singularities and formation of stable incoherent links in fluctuating speckled light. a**, Coherence characterization of fluctuating speckled light via the interference between different points and the central reference point $\boldsymbol{\rho}_0$. Zero degree of coherence ($|\mu(\boldsymbol{\rho}_b, \boldsymbol{\rho}_0)| = 0$) indicates a coherence singularity at $\boldsymbol{\rho}_b$. **b**, Incoherent Hopf links are composed of multiple coherence singularities in the speckled light. Blue and red distinguish the two links. In contrast, no stable intensity singularities can be observed in the instantaneous speckle light or its average intensity. **c,** Fluctuations in the intensity of the speckled light, averaging a uniform incoherent light field. **d**, Fluctuations in the amplitude and phase distribution.

To quantitatively analyze and design stable coherence singularities within an incoherent speckled light field, a mutual coherence function in the space-time domain $\Gamma(\boldsymbol{\rho}_1, \boldsymbol{\rho}_2, \tau) = \langle E^*(\boldsymbol{\rho}_1, t) E(\boldsymbol{\rho}_2, t+\tau) \rangle$ was introduced [40]. In this function, $E(\boldsymbol{\rho}, t)$ represents the instantaneous electric field, the asterisk signifies the complex conjugate, and the angular brackets indicate the ensemble average. In relatively steady fluctuation condition, this function can be transformed into a cross-spectral density function, represented by

$$W(\boldsymbol{\rho}_1, \boldsymbol{\rho}_2, \omega) = (1/2\pi) \int_{-\infty}^{\infty} \Gamma(\boldsymbol{\rho}_1, \boldsymbol{\rho}_2, \tau) e^{i\omega\tau} d\tau \qquad (1)$$

where $W(\boldsymbol{\rho}_1, \boldsymbol{\rho}_2, \omega) = \langle E^*(\boldsymbol{\rho}_1, \omega) E(\boldsymbol{\rho}_2, \omega) \rangle = \sum_n E_n^*(\boldsymbol{\rho}_1, \omega) E_n(\boldsymbol{\rho}_2, \omega)$. The frequency, denoted as $\omega$, is omitted from subsequent theoretical and experimental analyses as a result of the monochromatic light assumption. Consequently, Eq. (1) is simplified to a four-dimensional cross-spectral density function, $W(\boldsymbol{\rho}_1, \boldsymbol{\rho}_2)$, capable of characterizing and computing the second-order statistical characteristics of a propagating monochromatic and linearly polarized incoherent beam. The degree of coherence for a given pair of points $(\boldsymbol{\rho}_1, \boldsymbol{\rho}_2)$ within the incoherent field is defined as [40,41]:

$$\mu(\boldsymbol{\rho}_1, \boldsymbol{\rho}_2) = W(\boldsymbol{\rho}_1, \boldsymbol{\rho}_2)/\left[\sqrt{W(\boldsymbol{\rho}_1, \boldsymbol{\rho}_1)}\sqrt{W(\boldsymbol{\rho}_2, \boldsymbol{\rho}_2)}\right] \qquad (2)$$

Thus, point pairs with zero degree of coherence also exhibit zero cross-spectral density. At a specified value of $\boldsymbol{\rho}_2$, which is taken as a reference point, the four-dimensional coherence function simplifies to a two-dimensional (2D) function that remains complex-valued, containing both amplitude and phase components. Here, we designate the central point $\boldsymbol{\rho}_0$ as the reference point, and the 2D degree of coherence function $\mu$ can be expressed as:

$$\mu(\boldsymbol{\rho}, \boldsymbol{\rho}_0) = |\mu(\boldsymbol{\rho}, \boldsymbol{\rho}_0)| * \exp[i\varphi_\mu(\boldsymbol{\rho}, \boldsymbol{\rho}_0)] \qquad (3)$$

where $|\mu(\boldsymbol{\rho}, \boldsymbol{\rho}_0)|$ denotes the amplitude of the coherence function and $\varphi_\mu(\boldsymbol{\rho}, \boldsymbol{\rho}_0)$ represents its phase. Unlike conventional phase and polarization singularities which manifest in predetermined locations, the coherence structure and its coherence singularities exhibit variation depending on the chosen reference points. It is worth noting that, although the speckles fluctuate with time, the coherence structure of the light field remains stable. This statistical stability of the light field ensures invariant coherence singularities embedded within the coherence function, thereby maintaining the stability of incoherent knots and links. Our results show that while the amplitude of the coherence structure shifts with changes in the reference point, the central phase pattern of the coherence function remains almost unchanged, indicating the stability of the coherence singularity (see Fig. S1).

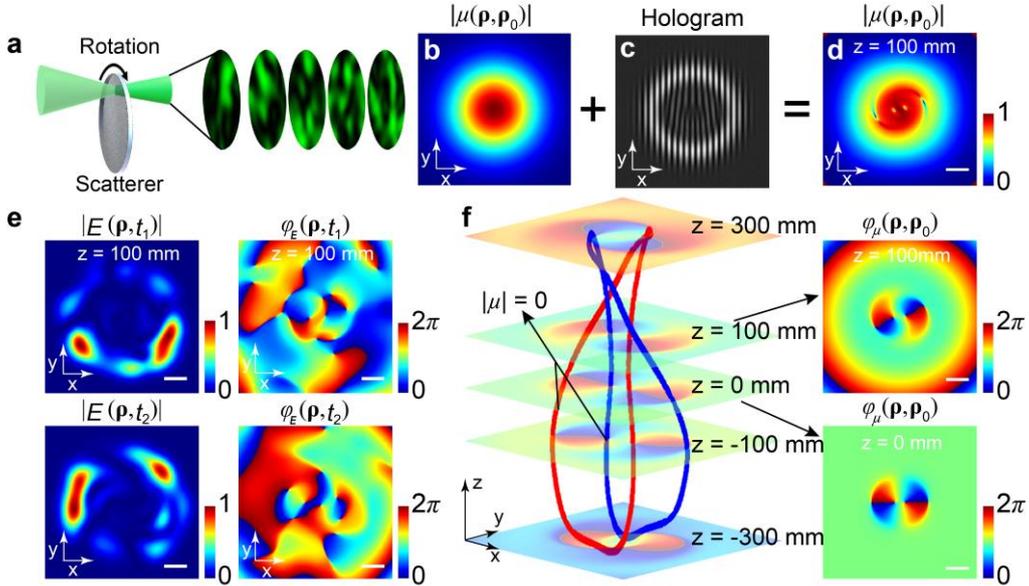

**Fig. 2 Numerical generation of incoherent light sources with fluctuating speckles and the formation of incoherent Hopf links. a,** Incoherent light source composed of speckles and **b,** its amplitude of coherence function. **c,** The hologram used to generate incoherent Hopf links. **d,** The amplitude of the coherence function after passing through the hologram, with zeros indicating coherence singularities. **e,** Corresponding instantaneous speckle amplitudes, and phase distributions after hologram. **f,** The incoherent Hopf link of coherence singularities and the typical phase structure of the coherence function on different planes. The scale bars represent 0.4 mm.

In Fig. 2a, an incoherent light source comprised of fluctuating speckle light fields was generated by the scattering of coherent light through a rough surface, and its amplitude of the coherence function exhibits a Gaussian distribution (Fig .2b). Afterwards, a hologram containing the vortex phase information was placed (Fig. 2c) [28]. When studying a 3D space following scattering, phase singularities emerge within a specific cross-sectional area, potentially leading to the formation of links and knots [42,43]. Nevertheless, the movement of the scatterer results in drastic and unpredictable changes to these phase singularities, links, and knots. After an extended period of statistical averaging, the fluctuation of the incoherent light field reaches a stationary distribution, which means its statistical properties do not change over time. The field generated by dynamic speckles may be considered an incoherent light field, in which the phase singularities wash out and vanish. By studying the coherence structure of the incoherent light field, it was observed that the 2D coherence structure changed from a Gaussian distribution (Fig. 2b) to an amplitude distribution with zero points (Fig. 1b), after the light passed through the hologram. Stable and orderly coherence singularities reemerged, although the instantaneous electric fields of the speckle light fields remained chaotic, with phase singularities appearing randomly (Fig. 2e). Based on this principle, a wider variety of coherence singularities can be designed and generated, such as topological links composed of coherence singularities.

In order to create an incoherent light field with a specific coherence structure, we used a computational hologram based on LG modal superposition (Fig. 2c), with a transmission function denoted as $H(\boldsymbol{\rho}) = H_{per}(\boldsymbol{\rho}) + H_{unper}(\boldsymbol{\rho}) = \sum C_n LG_p^l(\boldsymbol{\rho})$. Detailed coefficients $C_n$ are provided in Supplemental Table S1. The cross-spectral density after the hologram can be expressed as $W(\boldsymbol{\rho}_1, \boldsymbol{\rho}_2) = W_0(\boldsymbol{\rho}_1, \boldsymbol{\rho}_2) H^*(\boldsymbol{\rho}_1) H(\boldsymbol{\rho}_2)$, with $W_0(\boldsymbol{\rho}_1, \boldsymbol{\rho}_2)$ representing the incoherent light source with Gaussian distributed coherence structure. The cross-spectral density during propagation can be calculated based on extended Huygens-Fresnel integral (see Methods). The coherence amplitude generated is depicted in Fig. 2d, revealing two pairs of points with zero degrees of coherence ($|\mu| = 0$), referred to as coherence singularities. The phase of the coherence function $\varphi_\mu(\boldsymbol{\rho}, \boldsymbol{\rho}_0)$ is indeterminate at these points (Fig. 2f). Notably, $\varphi_\mu(\boldsymbol{\rho}, \boldsymbol{\rho}_0)$ is not the wavefront, as $\varphi_\mu$ represents the phase of coherence structure and is used to describe the spatial coherence between two points in space. Additionally, when analyzing structures for continuous propagation distances (Fig. 2f), points with zero degrees of coherence form an incoherent Hopf link, characterized by links with two nested rings. Consequently, a 3D topological structure emerges in the incoherent optical field during propagation.

In the experiment, the incoherent links were characterized using incoherent modal decomposition. The incoherent light field, considered as an incoherent superposition of multi-modes denoted as $W(\boldsymbol{\rho}_1, \boldsymbol{\rho}_2) = \sum_n E_n^*(\boldsymbol{\rho}_1) E_n(\boldsymbol{\rho}_2)$, was utilized. Following diffraction over a specific distance, the intensity $I_0(\mathbf{k})$ was treated as the incoherent superposition of diffraction patterns for multiple modes, represented by $I_0(\mathbf{k}) = \sum_n \psi_n^*(\mathbf{k}) \psi_n(\mathbf{k})$. Subsequently, the modes in $W(\boldsymbol{\rho}_1, \boldsymbol{\rho}_2)$ were updated using a multi-probe ptychography iterative engine based on a series of diffraction patterns. Utilizing modal superposition theory of spatial modes $\{E_n(\boldsymbol{\rho})\}$, one can reconstruct the cross-spectral density, intensity, and coherence function. The measurement is conducted on the source plane, which facilitates the derivation of spatial modes for other planes using angular spectral transmission based on the acquired spatial modes on the source plane. Furthermore, in the context of incoherent light sources with symmetrical coherence structures, the coherence length also serves as a metric of the degree of coherence, defined as the beam waist of the amplitude of the coherence function. The intensity correlations of the fluctuating speckles were used to characterize the coherence length (Supplemental Fig. S2).

Experimental results of incoherent links at $z = 100$ mm using a medium coherence light source (coherence length $\delta_0 = 1$ mm) are depicted in Fig. 3a, which corroborates the numerical simulations in Fig. 2d, f. The phase of the coherence structure over a range of propagation distances is calculated by applying angular spectral transmission on spatial modes of the source plane (Fig. 3b). By analyzing the phase distributions of the coherence function at propagation distances ranging from $z = -500$ mm to $z = 500$ mm, the locations of coherence singularities were tracked across 1001 evenly spaced planes and connected by lines. Fig. 3c visualizes the incoherent Hopf link composed of these coherence singularities. The experimental results align well with the those obtained from numerical simulations (see more details in Supplemental Fig. S3-4 and Fig. S7).

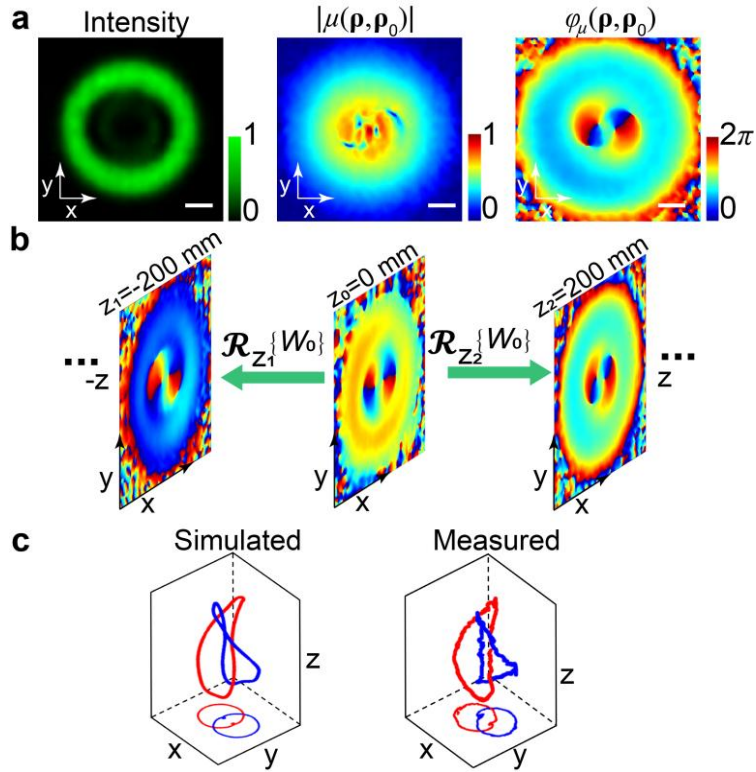

**Fig. 3 Experimental demonstration of incoherent links and 3D mapping of coherence singularities. a,** Experimental results of the beam intensity, and the amplitude and phase of coherence function at a plane $z = 100$ mm ($\delta_0 = 1$ mm). **b,** Experimental phase structures of the coherence function at various propagation planes reconstructed by using angular spectrum transmission. **c,** Numerical and experimental results of the incoherent links obtained by mapping coherence singularities in 3D plots. The scale bars represent 0.4 mm.

In the experimental configuration, a smaller spot incident on the dynamic scatterer produces a larger speckle size and a higher coherence of the light field. The coherence length can be adjusted by adjusting the size of the light spot on the scatterer. We chose three incoherent light sources with different degrees of coherence, each associated with different sizes of speckles. The results of high coherence ($\delta_0 = 2$ mm) are presented in Fig. 4a-d. Four coherence singularities emerge from the Gaussian-distributed coherence amplitude (Fig. 4b) after passing through the hologram and propagate to the plane $z = 100$ mm (Fig. 4 c,d). Conversely, the results of low coherence ($\delta_0 = 0.5$ mm) are shown in Fig. 4e-h, featuring a limited coherence amplitude envelope near the singularity (Fig. 4g). Regardless of high or low coherence, the phase structure of the coherence function remains consistent, and the positions of coherence singularities stay consistent (Fig. 3a, Fig. 4d and Fig. 4h). Various degrees of coherence were compared to assess the robustness of incoherent links, indicating the stability of these topological structures within incoherent light fields. However, considering the noise in the experiment, the coherence width is intentionally established to exceed the spatial extent of singularities. Failure to adhere to this precaution, which means the coherence is diminished too low, will result in a reduction of the coherence envelope. A decrease in the size of this envelope makes the coherence structure's reconstruction more vulnerable to noise.

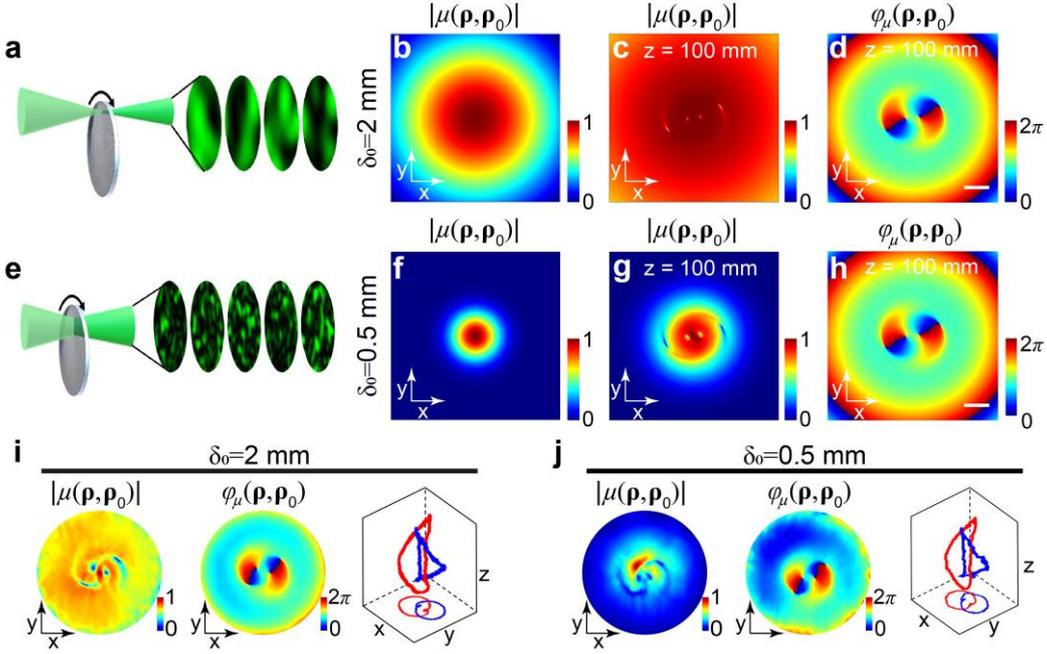

**Fig. 4 Numerical and experimental generation of incoherent links using light field of different coherence lengths. a,** Numerical incoherent light source composed of speckles under high coherence and **b,** its amplitude of the coherence function. **c-d,** Numerical amplitude and phase of the coherence function after passing through the hologram. **e,** Numerical incoherent light source under low coherence and **f,** its amplitude of the coherence function. **g-h,** Numerical amplitude and phase of the coherence function after passing through the hologram. Experimental results of incoherent Hopf links under **i**, high coherence and **j**, low coherence, including the amplitude and phase of the coherence function and incoherent links from 3D mapping of coherence singularities. The scale bars represent 0.4 mm.

Knots are also typical topological structures, and in this study, incoherent knots were effectively produced (Fig. 5a) by manipulating the distribution of coherence structure using the hologram shown in Fig. 5b. The corresponding LG mode coefficients are listed in Supplemental Table S1. The incoherent knots were also analyzed using multi-probe ptychography iterative engine, and angular spectral transmission was employed to reconstruct coherence structures during propagation, similar to the methodology described for incoherent links. Fig. 5a displays the reconstructed intensity, the amplitude of the coherence function, and the phase of the coherence function at the plane $z$ =100 mm for medium coherence ($\delta_0$ =1 mm), aligning with the results of the numerical simulation results in Supplemental Fig. S5. Tracking the coherence singularities with zero degree of coherence ($|\mu| = 0$), it is observed that their positions undergo gradual evolution during transmission. Different from the double-wire connection observed in incoherent links, these singularities are connected by a single wire (Fig. 5c-d). These knots are connected end to end by a single line with a crossing number of three, thus forming an incoherent Trefoil knot. The experimental results align well with numerical simulations. Fig. 5e illustrates incoherent knots of different degrees of coherence, which are consistent with the numerical simulations (Supplemental Fig. S7). Despite varying degrees of coherence for incoherent light fields, incoherent knots persist, demonstrating the robustness of incoherent knots against light field decoherence.

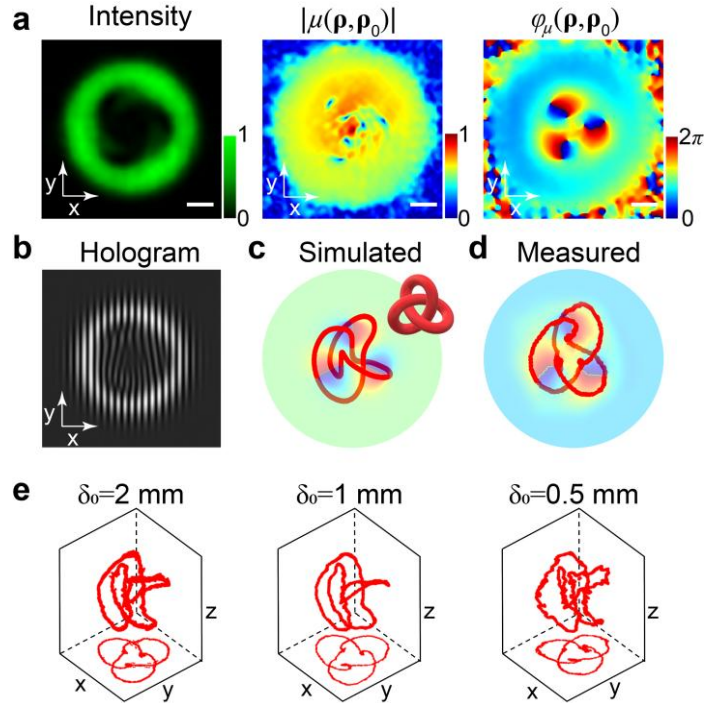

**Fig. 5 Experimental demonstration of incoherent Trefoil knots and comparison with numerical simulations. a**, Experimentally reconstructed beam intensity, and the amplitude and phase of coherence function at a plane $z$ =100 mm ($\delta_0$ =1 mm). **b,** The hologram used to generate incoherent Trefoil knots. **c-d**, Comparison of the numerical and experimental results for incoherent knots composed of coherence singularities. **e**, Experimental results of incoherent Trefoil knots reconstructed by mapping coherence singularities in 3D plots using different coherence lengths. The scale bars represent 0.4 mm.

**Discussions**

In this work, we have proposed and theoretically formulated the model of 3D topological structures in incoherent light fields - the incoherent links and knots. Remarkably, these 3D topological structures demonstrate robustness even in the presence of a reduced coherence and fluctuating light fields. We have devised a scheme to generate the 3D topological structures by inducing coherence singularities within an incoherent light field, thereby constructing both an incoherent Hopf link and an incoherent Trefoil knot. Further experimental measurements were achieved by incoherent modal decomposition, enabling the calculation of the coherence structure and the charting of the 3D topological configurations. Our experimental results confirm the successful generation and observation of incoherent links and knots in incoherent light fields.

A fully coherent structured light, when modulated by a specific hologram, can have its intensity zeros (phase singularities) evolve and form a 3D topological structure. However, with a fluctuating incoherent light source and the same hologram, the light field fluctuations cause the singularities to wash out and eventually vanish (Fig. S8). The decoherence of light source makes the identification of singularities and corresponding links and knots infeasible. In contrast, the coherence singularities, corresponding to zero degrees of coherence, remain robust as a 3D topological structure. In addition, according to Eq. (1), the cross-spectral density function is proportional to the electric field as the degree of coherence approaches infinity and there are no fluctuations in spatial modes [44]. Thus, the 3D topological structure constructed by phase singularities in a coherent light field could be viewed as a special case within the broader class of 3D topological structures framed by coherence singularities.

For the topological structures, we have also investigated the robustness of incoherent links and knots against different perturbations, considering factors such as phase perturbation to the hologram and fluctuation of speckle noise. Fig. S9 illustrates the impact of phase noise on coherence singularities. Panels S9a-c show the phase perturbation ($C_n^2 = 10^{-10}$) and the corresponding incoherent link. As long as the noise level remains within a certain range, the coherence singularities that define these structures still form the same topological links or knots, even though they may shift or deform slightly. However, with further increasing the noise level or introducing phase singularities into the noise [45], coherence singularities may appear and disappear, potentially undergoing a topological transition. In Fig. S10, we further examine the effect of adding more randomly fluctuating speckles to the detection plane. By considering the incoherent superposition of a structured light field with noise of varying energy levels, we define their ratio as the signal-to-noise ratio (SNR). In Fig. S10, when the SNR is greater than 1:1, the evolution of coherence singularities observed from phase patterns of the coherence structure remains stable, even though the recognition accuracy of singularities in amplitude is very low. As the noise level further increases, however, the coherence singularities are progressively destroyed, eventually disrupting the link they form.

It is expected that further research into topological transitions or knot invariants mediated by coherence singularities would emerge in the future. For example, new findings may lead to encoding high-dimensional optical information using incoherent links and knots, where each distinct topological structure can represent different data values [33]. A continuous transition from a link to a knot has been studied by manipulating the topological charge of the LG mode [46]. The ability of incoherent topological structures to encode information by mapping various topological configurations to data sets could offer new possibilities for data transmission. Moreover, leveraging the robustness against noise and fluctuations [45], information encoding based on incoherent topological structures could be robust against perturbations, potentially enhancing both the security and efficiency for optical communication systems [47, 48] without specially designed topological vortex guides [49].

In summary, incoherent links and knots exist in the coherence structures of incoherent light fields, that is, second-order correlations. Different from polarization and phase singularities, which also form 3D topological structures, coherence singularities are higher-dimensional topological information of the light field. Our results of incoherent knots and links stem from the analysis of the fluctuating light field's statistical properties. The discovery of such incoherent topological structures provides a new degree of freedom in the generation of 3D topological structures. Although our results are obtained in an incoherent light field composed of dynamic speckle fields, the underlying physics with incoherent sources can encompass pseudo-thermal light, LEDs, X-rays, or even electron beams. Moreover, these results provide new insights into the evolution and manipulation of coherence singularities that may have potential applications in information coding and quantum optics.

## Materials and methods

**Theoretical generation of incoherent links and knots in incoherent light fields.** The incident quasi-monochromatic random fluctuations of the incoherent beam are characterized by the cross-spectral density function in space-frequency, denoted as: $W(\boldsymbol{\rho}_1, \boldsymbol{\rho}_2, 0) = \langle E^*(\boldsymbol{\rho}_1) E(\boldsymbol{\rho}_2) \rangle$. Following modulation by a spatial light modulator (SLM) and propagation over a distance z, the cross-spectral density function is transformed to:

$$W(\mathbf{r}_1, \mathbf{r}_2, z) = \left(\frac{k}{2\pi z}\right)^2 \iint W(\boldsymbol{\rho}_1, \boldsymbol{\rho}_2, 0) H^*(\boldsymbol{\rho}_1) H(\boldsymbol{\rho}_2)$$

$$\times \exp\left[-\frac{ik}{2z}(\mathbf{r}_1 - \boldsymbol{\rho}_1)^2 + \frac{ik}{2z}(\mathbf{r}_2 - \boldsymbol{\rho}_2)^2\right] d\boldsymbol{\rho}_1 d\boldsymbol{\rho}_2 \quad (4)$$

$H(\boldsymbol{\rho})$ represents the transmittance function of the SLM, **r** signifies the coordinates at transmission distance z, and $k$ denotes the wavevector. The distribution of coherence structures in the z-plane can be determined using Eq. (2) presented in the main text. In order to create incoherent links or knots, it is necessary for the cross-spectral density of the source plane to adhere to a particular distribution. By manipulating the parameters of the incoherent light field within the source plane, a coherence structure with a specific distribution can be produced. The accompanying supplement offers further theoretical information, such as the generation of speckle in incoherent optical fields, SLM transmittance functions, and knot theory.

**Experimental generation of incoherent links and knots via coherence singularities.** As illustrated in Supplemental Fig. S11a, the experimental configuration for the generation and measurement of incoherent links and knots is outlined. A coherent laser beam is generated by a semiconductor laser (VentusHR 532 nm) and magnified by a beam expander. Subsequently, the magnified beam is directed onto a ground glass surface (which may be substituted with another scattering medium) by a thin lens L1 ($f$ = 100 mm). Following collimation by a thin lens L2 ($f$ = 150 mm), the beam is transformed into a speckle pattern at the rear focal plane of L2, designated as the source plane. The size of the speckle is influenced by both the focal spot size on the ground glass and the roughness of the ground glass. In our case, the requirement that the beam spot diameter on the scattered sample exceeds the non-uniform scale of the ground glass is met. As the ground glass rotates, these speckles exhibit random fluctuations and are superimposed incoherently to generate an incoherent light field. The autocorrelation of these speckles can be used to calculate the coherence length, as illustrated in Supplemental Fig. S2. As the focal spot size increases, the size of source speckles decreases and the source coherence length decreases. Following the source plane, an SLM (HDSLM85T, pixel pitch=8 μm) loaded with holograms $H(\boldsymbol{\rho})$ (Fig. 2c and Fig. 5b) is utilized to modulate the amplitude and phase of all instantaneous speckle fields, thereby altering the coherence structures of the incoherent light field. The modulated incoherent beam is then transmitted through a 4$f$ system consisting of lenses L3 and L4, ultimately imaging onto the back focus plane of L4 to achieve the desired coherence structures of incoherent Hopf links and Trefoil knots.

**Acknowledgements**

This work was supported by the National Key Research and Development Program of China (No. 2022YFA1404800), National Natural Science Foundation of China (No. 12174280, No. 12204340, No. 12192254, No. 92250304, No. 12434012, No. W2441005), Priority Academic Program Development of Jiangsu Higher Education Institutions, and Postgraduate Research & Practice Innovation Program of Jiangsu Province (KYCX24_3287).

**Data availability**

All the data supporting the findings of this study are available within this article and its Supplementary information. Any additional information can be obtained from corresponding authors on reasonable request. Source Data are provided with this paper.

**Conflict of interests**

The authors declare no competing interests.

**Contributions**

Z.W., X.L., C.Z and Y.C. proposed the original idea and performed the theoretical analysis. Z.W.

and X.L performed the experiments, data analysis and contributed to development the measurement method. C.Z., Y.C and Z.C. supervised the project. All the authors contributed to the data analysis and writing of the manuscript.

**References**

1. Adams, C. C. The Knot Book: An Elementary Introduction to the Mathematical Theory of Knots. (Providence: American Mathematical Society, 1994).
2. Kauffman, L. H. Knots and Physics. 3rd ed. (Singapore: World Scientific, 2001).
3. Davies, A. et al. Advancing mathematics by guiding human intuition with AI. *Nature* **600**, 70-74 (2021).
4. Irvine, W. T. M. & Bouwmeester, D. Linked and knotted beams of light. *Nature Physics* **4**, 716-720 (2008).
5. Witten, E. Quantum field theory and the Jones polynomial. *Communications in Mathematical Physics* **121**, 351-399 (1989).
6. Wang, J. W. et al. Experimental observation of Berry phases in optical Möbius-strip microcavities. *Nature Photonics* **17**, 120-125 (2023).
7. Moffatt, H. K. The degree of knottedness of tangled vortex lines. *Journal of Fluid Mechanics* **35**, 117-129 (1969).
8. Zhang, Q. X. et al. Ferromagnetic switching of knotted vector fields in liquid crystal colloids. *Physical Review Letters* **115**, 097802 (2015).
9. Tai, J. S. B. & Smalyukh, I. I. Three-dimensional crystals of adaptive knots. *Science* **365**, 1449-1453 (2019).
10. Tai, J. S. B., Wu, J. S. & Smalyukh, I. I. Geometric transformation and three-dimensional hopping of Hopf solitons. *Nature Communications* **13**, 2986 (2022).
11. Berger, M. A. Introduction to magnetic helicity. *Plasma Physics and Controlled Fusion* **41**, B167 (1999).
12. Poy, G. et al. Interaction and co-assembly of optical and topological solitons. *Nature Photonics* **16**, 454-461 (2022).
13. Zhang, H. K. et al. Creation of acoustic vortex knots. *Nature Communications* **11**, 3956 (2020).
14. Zhao, H. Q. et al. Liquid crystal defect structures with Möbius strip topology. *Nature Physics* **19**, 451-459 (2023).
15. Kleckner, D. & Irvine, W. T. M. Creation and dynamics of knotted vortices. *Nature Physics* **9**, 253-258 (2013).
16. Wan, C. H. et al. Scalar optical hopfions. *eLight*, **2**, 22 (2022).
17. Zhong, J. Z., Wan, C. H. & Zhan, Q. W. Optical twisted phase strips. *ACS Photonics* **10**, 3384-3389 (2023).
18. Veretenov, N. A., Fedorov, S. V. & Rosanov, N. N. Topological vortex and knotted dissipative optical 3D solitons generated by 2D vortex solitons. *Physical Review Letters* **119**, 263901 (2017).
19. Guo, X. et al. Tying polarization‐switchable optical vortex knots and links via holographic all‐dielectric metasurfaces. *Laser & Photonics Reviews* **14**, 1900366 (2020).
20. Larocque, H. et al. Optical framed knots as information carriers. *Nature*

*Communications* **11**, 5119 (2020).

21. Wang, X. L. et al. Manipulating propagation and evolution of polarization singularities in composite Bessel-like fields. *Photonics Research* **11**, 121 (2023).

22. Berry, M. V. & Dennis, M. R. Knotted and linked phase singularities in monochromatic waves. *Proceedings of the Royal Society of London. Series A: Mathematical, Physical and Engineering Sciences* **457**, 2251-2263 (2001).

23. Leach, J. et al. Knotted threads of darkness. *Nature* **432**, 165 (2004).

24. Dennis, M. R. et al. Isolated optical vortex knots. *Nature Physics* **6**, 118-121 (2010).

25. Leach, J. et al. Vortex knots in light. *New Journal of Physics* **7**, 55 (2005).

26. Romero, J. et al. Entangled optical vortex links. *Physical Review Letters* **106**, 100407 (2011).

27. Bauer T. et al. Observation of optical polarization Möbius strips. *Science* **347**, 964-966 (2015).

28. Larocque H. et al. Reconstructing the topology of optical polarization knots. *Nature Physics* **14**, 1079-1082 (2018).

29. Pisanty E. et al. Knotting fractional-order knots with the polarization state of light. *Nature Photonics* **13**, 569-574 (2019).

30. Forbes, A. et al. Structured light. *Nature Photonics* **15**, 253-262 (2021).

31. Wan, C. H., Chong, A. & Zhan, Q. W. Optical spatiotemporal vortices. *eLight*, **3**, 11 (2023).

32. Li P. et al. Optical vortex knots and links via holographic metasurfaces. *Advances in Physics: X* **6**, 1843535 (2021).

33. Kong L. J. et al. High capacity topological coding based on nested vortex knots and links. *Nature Communications* **13**, 2705 (2022).

34. Kong, L. J. et al. Topological holography and storage with optical knots and links. *Laser & Photonics Reviews* **17**, 2300005 (2023).

35. van Dijk, T., Schouten, H. F. & Visser, T. D. Coherence singularities in the field generated by partially coherent sources. *Physical Review A* **79**, 033805 (2009).

36. Palacios, D. M. et al. Spatial correlation singularity of a vortex field. *Physical Review Letters* **92**, 143905 (2004).

37. Lu, X. Y. et al. Phase detection of coherence singularities and determination of the topological charge of a partially coherent vortex beam. *Applied Physics Letters* **114**, 201106 (2019).

38. Rana A. et al. Potential of attosecond coherent diffractive imaging. *Physical Review Letters* **125**, 086101 (2020).

39. Lu, X. Y. et al. Four-dimensional experimental characterization of partially coherent light using incoherent modal decomposition. *Nanophotonics* **12**, 3463-3470 (2023).

40. Wolf, E. Introduction to the Theory of Coherence and Polarization of Light (Cambridge: Cambridge University Press, 2007).

41. Goodman, J. W. Statistical Optics. 2nd ed. (Hoboken: John Wiley & Sons, 2015).

42. O'holleran, K. et al. Fractality of light's darkness. *Physical Review Letters* **100**, 053902 (2008).

43. O'holleran, K., Dennis, M. R. & Padgett, M. J. Topology of light's darkness. *Physical Review Letters* **102**, 143902 (2009).

44. Wolf, E. New theory of partial coherence in the space–frequency domain. Part I: spectra and cross spectra of steady-state sources. *Journal of the Optical Society of America* **72**, 343-351 (1982).

45. Pires, D. G. et al. Knots of darkness in atmospheric turbulence. Print at https://doi.org/10.48550/arXiv:2401.12306 (2024).

46. Zhong, J. Z. et al. Observation of optical vortex knots and links associated with topological charge. *Optics Express* **29**, 38849-38857 (2021).

47. Peng, D. M. et al. Optical coherence encryption with structured random light. *PhotoniX* **2**, 6 (2021).

48. Cheng, M. J. et al. Metrology with a twist: probing and sensing with vortex light. *Light: Science & Applications* **14**, 4 (2025).

49. Hu, Z. C. et al. Topological orbital angular momentum extraction and twofold protection of vortex transport. *Nature Photonics* **19**, 162-169 (2025).